\documentstyle[prl,aps]{revtex}                
\input{epsf}
\draft
\begin{document}

\title{Broken-Bond Rule  for the Surface  Energies of Noble Metals}
\author{I. Galanakis, G. Bihlmayer, V. Bellini, N. Papanikolaou, R. Zeller,
S. Bl\"ugel,\cite{SBaddress} and P. H. Dederichs}

\address{Institut f\"ur Festk\"orperforschung, Forschungszentrum J\"ulich, 
D-52425 J\"ulich, Germany}

\date{\today}
\maketitle

\begin{abstract}
Using two different full-potential {\em ab initio} techniques       
we introduce a simple, universal rule based on the number of         
broken first-neighbor bonds to determine                             
the surface energies of the three noble metals Cu, Ag and Au.
When a bond is broken, the rearrangement of the electronic charge   
for these metals does not lead to a change of the remaining bonds.
Thus  the energy needed to break a bond is independent of the surface
orientation. This novel finding can lead to the development of
simple models to describe the energetics of a surface               
like  step and kink formation,  crystal growth, 
alloy formation,  equilibrium                                
shape of mesoscopic crystallites and surface faceting.
\end{abstract}


The surface energy represents a fundamental material property.
It is given by half the energy needed to cut a 
given crystal into two half crystals. As such the surface energy 
naturally depends on the strength of the bonding and on the orientation of 
the surface plane.  
A variety of experimental techniques have been developed to measure the surface energy 
\cite{kumikov}, but all measurements are performed at high temperatures where surfaces
are badly defined.  The most comprehensive                             
 data  stem from surface tension measurements in the liquid 
phase and by extrapolating the resulting orientation-averaged surface free energies 
to zero temperature
\cite{Tyson,Boer}. 
The knowledge of the orientation-dependence of the 
surface energies is necessary to predict the  
equilibrium shape of a mesoscopic crystal and  
to study a series of important phenomena in materials science
like  crystal growth,
creation of steps and kinks on surfaces, growth, stability and alloy formation of 
thin films or surface-melting faceting.

The lack of experimental data can be replaced by {\em ab initio}
calculations. Due to the development of the density functional theory
during the last two decades, {\em ab initio} methods are able to calculate
many physical properties with unprecedented  accuracy.
Methfessel and collaborators \cite{Methfessel} were the first to study the trends in
 surface energy, work function and relaxation for the whole
series of bcc and fcc 4$d$ transition metals, using a
full-potential (FP) version of the linear muffin-tin orbital (LMTO)
method in conjunction with the local density approximation (LDA)
to the exchange-correlation potential. In the
same spirit Skriver and co-workers have used a  LMTO technique to
calculate the surface energy and the work function of most of the
elemental metals \cite{SkriverPRB92}. Recently, 
Vitos and collaborators using their full-charge Green's function
LMTO technique   in the atomic sphere approximation (ASA) in
conjunction with the generalized gradient approximation (GGA)
elaborated a very  useful  database that contains the low-index surface
energies for 60 metals in the periodic table \cite{VitosSurf98}.
Also many semi-empirical \cite{semi} and
tight-binding \cite{tb} studies exist. 

In this contribution we show, using precise  {\em ab initio} techniques, that 
irrespective of the orientation the      
surface energies of Cu, Ag, and Au are simply proportional to
 the number of broken bonds between  a surface atom and its nearest neighbors;
for all surface orientations, except the (111) and (100), one has to  take into
account in the total number of broken bonds also the nearest bonds lost
by the subsurface atoms.                                                 
We demonstrate this in
 calculations for the low-index surfaces (111), (100), and (110) as well as for four     
vicinal surfaces. The resulting anisotropy ratios,  i.e.  the ratio of 
the surface energy for a given surface orientation 
with respect to the (111) surface energy, practically 
always agree with the 
``ideal'' broken bond ratios, i.e.\ the number of  broken bonds between nearest 
neighbors for this surface  with respect to the (111) surface. 
This novel finding implies that  (i) for the          
noble-metal surfaces the interaction between an atom and its second and 
further neighbors is very
small and, that (ii) the charge rearrangement caused by the bond breaking does  
not significantly change the strength of the remaining bonds. 
Therefore, the energy needed to break a bond does not depend                            
on the orientation, so that
for each noble metal the surface energy for only one orientation is needed. 
We show that highly accurate calculations are compulsory  to obtain these 
results.
 
 To perform the calculations, we have used both the
full-potential screened Korringa-Kohn-Rostoker (FKKR) method
\cite{Zeller95}, and the full-potential linearized
augmented plane wave (FLAPW) method \cite{Wimm:81} as implemented in
the FLEUR code in conjunction with LDA.
 All calculations have been performed using  the experimental lattice
parameters: 3.61 \AA\ for Cu, 4.09 \AA\ for Ag, and 4.08
\AA\ for Au.   
For the FKKR an angular momentum cut-off of 
$\ell_{max}$=3 for the wavefunctions and of 
$\ell_{max}$=6 for the multipole expansion of the charge density
and the potential has been used.
To calculate the charge density, we integrated the Green's function
along a contour on the complex energy plane, which extends from
the bottom of the valence band up to the Fermi level \cite{Zeller82}. Due
to the smooth behavior of the Green's functions for complex
energies, only few energy points are needed;  27 points have 
been used. 
A very large number of  {\bf k}$_\parallel$ points 
in the irreducible part of the two-dimensional Brillouin zone (2D-IBZ) has been used for the decisive  
complex  energies close to the Fermi level 
($\sim$ 300 points for the (111) and up to $\sim$                           
800 points for the (110) surface).
The surface energy $\gamma$ for a $N$-layer slab embedded in semi-infinite
vacuum is given by
$\gamma=(E_{s}-N E_{b})/2$,                        
where $E_{s}$ is the total energy of the slab     
and $E_{b}$ is the  energy per atom in the bulk crystal; 
note that for fcc crystals $N$ is also the number of inequivalent atoms
in the slab. To be consistent for    
all the cases we used as $E_{b}$ the energy per atom of the                      
central layer of the slab. We have converged the number of metal 
layers so that the surface energies are converged within 0.01 eV;
we used 12 layers for the (111), 14 for
the (100), and 18 for the (110) surface. 
In FLAPW we calculated the surface energies
from the total energies of two films of different thickness:
for the (111) surface 13- and 11-layer films were
used, for the (100) orientation 15 and 13 layers and for the (110)
surfaces 19- and 17-layer films were calculated. A basis set of 80-90 augmented 
plane waves per atom and 425, 325, and 408 {\bf k}$_\parallel$ points were used 
in the 2D-IBZ of the (111), (100), and (110) surface, respectively.

In Table~\ref{table1} we collect the scalar-relativistically calculated surface
energies within both  FKKR and FLAPW and we compare them with the
values obtained by the LMTO in Ref.~\onlinecite{VitosSurf98}.
For Cu and Ag, the absolute values calculated with all three methods
agree nicely.  For Au both FKKR and LMTO predict 
that the surface energies are very close to the Cu values, whereas FLAPW
predicts similar surface energies for Au and Ag in agreement with previous
FLAPW results, 0.66 eV for Ag(100) and
0.67 eV for Au(100) \cite{eibler}. A pseudopotential technique 
on the other hand 
shows the FKKR behavior:
$\gamma$= 0.58 eV for 
Ag(100) and 0.70 eV for  Au(100) \cite{Ho}.

\begin{figure}\begin{center}
\begin{minipage}{3.0in}
 \epsfxsize=3.0in \epsfysize=2.5in \centerline{\epsfbox{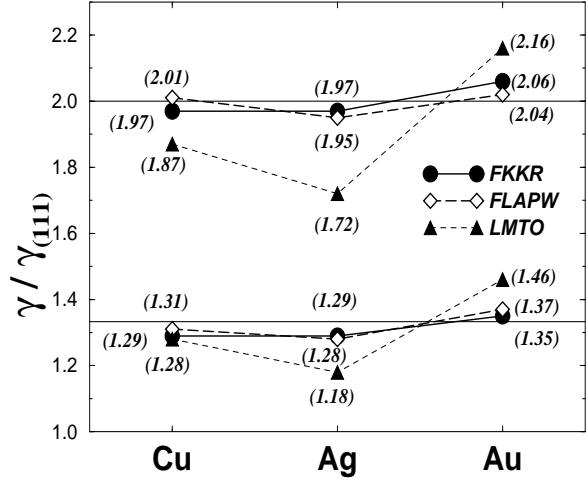}}
 \end{minipage}\end{center}
 \caption{ The anisotropy ratios for the three noble metals,
           $\gamma_{(100)}/\gamma_{(111)}$ and
           $\gamma_{(110)}/\gamma_{(111)}$,
   using both FKKR and FLAPW. The
   LMTO results are from Ref.~{\protect\onlinecite{VitosSurf98}}.
FP-LMTO calculations  {\protect\onlinecite{Methfessel}}
 for Ag produced similar results
 to those  in  Ref.~{\protect\onlinecite{VitosSurf98}}.
Surface energies are    calculated in eV/(surface atom). The
   two straight solid lines represent the ideal first neighbor
   broken-bond ratios; 4/3 for (100) and
   6/3=2 for (110).} \label{fig1}
\end{figure}

As central result we present in Fig.~\ref{fig1}                                  
the anisotropy ratios. We remark that both  
FKKR and FLAPW calculations produce practically the
same anisotropy ratios for all the noble metals, while previous
LMTO \cite{VitosSurf98} and FP-LMTO  \cite{Methfessel} calculations
gave anisotropy ratios that deviate considerably from the present 
results, especially for Ag.
Notably, both FKKR and FLAPW give results that are very close                  
to 4/3 for the $\gamma_{(100)}/\gamma_{(111)}$ ratio and close to 6/3
for the $\gamma_{(110)}/\gamma_{(111)}$ ratio. These are
exactly the ratios between the number of first-neighbor broken   
bonds for these surfaces. 
 This finding can lead to two independent conclusions.
Firstly, the broken bonds between a surface or a subsurface atom and its second and further
neighbors have a negligible contribution to the surface energy.
Secondly, the energy needed to break
a bond is the same for any surface orientation. This is surprising,  
since one expects that breaking a bond in a                          
surface leads to a rearrangement of the electronic charge
resulting in a  strengthening of the remaining bonds, so
that one needs more energy to break them. But it seems that this bond
strengthening, due to the reduction of neighbors, is negligible for  the noble metals.
To examine whether  this finding  also holds for the vicinal surfaces 
we used the FKKR method and calculated the surface energies and the anisotropy ratios
for the next four more close-packed surfaces (see Table~\ref{table2}).
The number of layers used in the calculations is
21 for the (113), 30  for the (331) and the (210), and
finally 32 for the (112) surface.   
For all these surfaces the anisotropy ratios are close to the 
ratios given by the broken first-neighbor bonds. Au shows slightly larger
deviations from these ideal ratios compared to Ag and Cu, which attends 3.6\%\ in the
case of the (210) surface.
So the free energy  $\gamma_{(hkl)}$ in eV/(surface atom) needed to create any surface with
a Miller index $(hkl)$ reduces just to the product  of                                   
$\gamma_{(111)}$ and the ratio of the first-neighbor broken 
bonds, $N_{(hkl)}$, and $N_{(111)} = 3$:
\begin{equation}
 \gamma_{(hkl)} = \frac{N_{(hkl)}}{3} \gamma_{(111)}.
\end{equation}
$N_{(hkl)}$ can be easily obtained for any fcc surface~\cite{mackenzie}:
\begin{equation}
N_{(hkl)} = \left\{ \begin{array}{ll} 2h+k & \mbox{$h,k,l$ odd} \\
4h+2k& \mbox{otherwise} \end{array} \right. \qquad h \ge k \ge l.
\end{equation}

Given the disagreement of our anisotropy ratios with the published results of
\cite{Methfessel,VitosSurf98}, we have performed extensive tests regarding the
accuracy and consistency of our calculations. With the FKKR method we have   
performed calculations also for the somewhat smaller
LDA lattice constants. Moreover, we have used both the full-potential KKR code
as well as the KKR code with ASA potentials and full-charge density. 
While the absolute values of the surface energies change, the anisotropy
ratios are extremely stable. With the FLAPW code we have performed additional 
calculations with the GGA and have estimated the effect of spin-orbit coupling for Au. 
Whereas the latter tends to increase the absolute values of the surface energies,      
GGA lowers the absolute values, but again                                                
we obtained anisotropy ratios in very good agreement 
with the broken-bond rule.

\begin{figure}
\begin{center}
\begin{minipage}{2.3in}
 \epsfxsize=2.5in \epsfysize=2.3in \centerline{\epsfbox{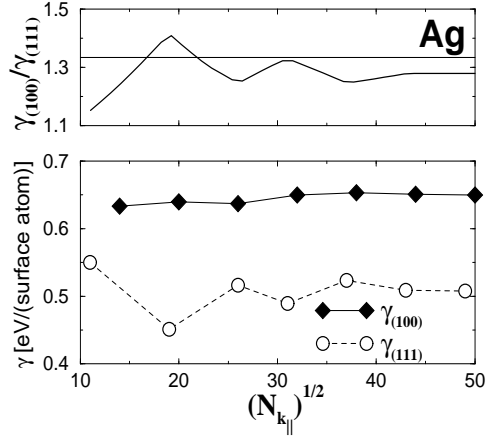}}
 \end{minipage}
\end{center}
 \caption{ Convergence of the surface energies $\gamma{(111)}$ and $\gamma{(100)}$  with th
e
square root of   the number of {\bf k}$_\parallel$-points in the full first
2D Brillouin zone using the FLAPW method.  The upper panel shows the resulting
anisotropy ratio  $\gamma_{(100)}/\gamma_{(111)}$ together with the ideal value of 4/3.
   \label{fig2}}
  \end{figure}

In the course of these checks we identified the most probable reason 
for the failure of previous calculations. 
In Fig.~\ref{fig2} we present the convergence of the
surface energy of Ag with respect to the square root of the number 
of {\bf k}$_\parallel$-points
used to perform integrations in the full first Brillouin zone,
as for this system the difference between our present results and the
ones in Refs.~\cite{Methfessel} and \cite{VitosSurf98} 
is the largest. The (111) surface energy 
is very sensitive to the number of {\bf
k}$_\parallel$-points, while this is not the case for the (100) surface; the
(110) surface behaves similar to the (100) surface.
This sensitivity might arise from 
 a surface state centered at the $\bar{\Gamma}$ point, which all three noble metals 
possess and 
which requires a very dense {\bf k}$_\parallel$-grid to
account for it. This is in-line with the observation that 
the largest deviations are obtained for Ag, for which this
surface state is closest to the Fermi level. In the upper panel of Fig.~\ref{fig2} 
we show the ratio between the (100) and the (111) surface energies, which
follows the oscillations of the $\gamma_{(111)}$ energy. In
Refs.~\cite{Methfessel} and \cite{VitosSurf98} the square
root of the number of {\bf k}$_\parallel$-points in the full Brillouin 
zone used is about 9
and 13, respectively. For these numbers of {\bf k}$_\parallel$-points 
the FLAPW method gives a ratio which
is between 1.15-1.20 very close to the ratio of 1.18 obtained by
both LMTO methods. Thus the non-convergence of the surface energies with the number 
of {\bf k}$_\parallel$-points is the reason for the 
differences seen in Fig.~\ref{fig1}.

In the following we discuss the range of validity and the limits of the broken-bond 
rule. Firstly, we note that we obtain similar
anisotropy results for the anisotropy ratios of the Rh, Ir, Pd, and Pt surfaces,
although with slightly larger deviations (about $\pm$3-5\%) from the ideal ratios~\cite{Gala:01}.           
On the other hand, if we compare with another broken-bond system,
i.e.\ the single vacancy, we find no agreement at all. In this case each of the 12 nearest   
neighbors of the vacancy looses one bond, so that according to the above rule 
the vacancy formation energy should be 12/3 times the (111) surface energy. In reality,
however, it is for all three noble metals more than a factor two smaller. 
Analogously we expect that the cohesive energy is by the 
same factor 12/3 larger than the (111) surface energy. The values estimated in this 
way for the cohesive energies of Cu, Ag, and Au (2.70 eV, 2.26 eV, and 2.49 eV,
respectively) are about 20\%-25\% smaller than the experimental data 
(3.51 eV, 2.99 eV, and 3.56 eV, respectively)~\cite{kittel}.

Finally we studied the effect of lattice relaxations on the calculated surface energies 
using the FLAPW and allowing the three first layers to relax. 
Although the calculated surface energies
change, the effect on the anisotropy ratios is much smaller. For Cu, relaxations 
reduce $\gamma$ by 0.8\%, 0.7\% and 2.6\% for the (111), (100) and (110) surfaces,
respectively. The (100)/(111) anisotropy ratio stayed unchanged by the relaxations
while the relaxed (110)/(111) ratio was 1.98 compared to the value of 2.01 for 
the unrelaxed structure. For Ag, the surface energy was reduced by 0.1\%, 0.7\% and
1.8\% for the three low-index surfaces, respectively. The new anisotropy ratios are 
1.27 and 1.93 slightly smaller than the original values of 1.28 and 1.95.
Finally for Au, relaxations reduce the surface energies by 0.2\% and 0.8\% for the 
(111) and (100) surfaces and the anisotropy ratio becomes 1.35 slightly smaller than
the value, 1.37, for  the unrelaxed structure. The Au(110) surface shows a large relaxation;
the distance between the first and the second layer ($\Delta d_{12}$) is 
reduced by 13.8\%, the $\Delta d_{23}$ is expanded by 6.9\% and finally the 
$\Delta d_{34}$ is also reduced by 3.2\%. The surface energy is reduced by 6.5\% and the 
anisotropy ratio is now 1.89 compared to the 2.04 for the unrelaxed structure, but remains 
close to the broken-bond rule value of 2.0. So even large relaxations have a rather small 
impact on the calculated anisotropy ratios which are reasonably described by the 
broken-bond rule.

Several experiments at high 
temperatures have been carried out mainly on gold crystallites  \cite{gold-cryst}
 to determine  the $\gamma$-anisotropy, but their interpretation is difficult.  
Entropy terms, describing 
 the lower vibrational frequencies of the atoms at the surface as compared to
the bulk, the formation of kinks and finally the creation of holes and pillboxes at the low-index surfaces,
have to be added to the total free energy. Also at such high temperatures 
 the  surface-melting faceting \cite{frenken}, i.e.  the break-down 
of a vicinal surface in a dry and a
melted one, plays a predominant role. 
Recently, Bonzel and Edmundts  \cite{bonzel} have shown that analyzing the equilibrium
shape of crystallites at various temperatures by 
scanning tunneling microscopy can yield absolute values of
 the surface and step energies versus temperature,
but this technique has not been yet applied.
Also after the growth of 
a nanocrystal, there is a dynamical procedure towards its equilibrium, 
which can be modeled using  the anisotropy of the surface energies at 0 K \cite{combe}.
Similar problems arise from high-temperature measurements on single Au surfaces 
\cite{breuer}. The easy calculation of the surface energies 
for the noble metals as a product of the $\gamma_{(111)}$ and the number of first 
neighbor broken bonds
can be used to develop more complicated models describing the above phenomena. 
There are also other important applications in materials science like the growth 
and stability of a thin film or the growth of a crystal surface where 
the knowledge of the absolute values and the anisotropy of the surface energies is 
crucial.

In this Letter we have shown for different surface orientations 
that the surface energies of the noble metals 
accurately scale with the number
of the  broken nearest-neighbor bonds, so that the calculated anisotropy ratios
 always agree well with the ideal broken-bond ratios. We have demonstrated this 
in FKKR and FLAPW calculations for seven low-index and vicinal surfaces of
Cu, Ag and Au.
We believe that the simplicity of these results are of great interest for a  
variety of problems in materials science like step, kink, and alloy formation,              
crystal growth, surface-melting faceting or the shape of small crystallites on a 
catalyst.

Authors wish to thank H.P. Bonzel for helpful comments
on the manuscript. Authors  gratefully acknowledge support from the {\em Psi-k}
 TMR network
(Contract No: FMRX-CT-0178), the TMR network of {\em Interface Magnetism}
(Contract No: ERBFMRXCT96-0089) and the RT Network of {\em Computational
Magnetoelectronics} (Contract No: RTN1-1999-00145) of the European Commission.

\begin{table}
\caption{Scalar-relativistic surface energies  for the three noble
metals using both FKKR and FLAPW 
compared with previous LMTO results from             
Ref.~{\protect\cite{VitosSurf98}}. All results are given in
  eV/(surface atom).}
\label{table1}
\begin{tabular}{c|ccc|ccc|ccc}
\multicolumn{1}{c}{} &  \multicolumn{3}{c}{Cu} &
\multicolumn{3}{c}{Ag} & \multicolumn{3}{c}{Au} \\ {\small
$\gamma$ (eV)} & {\tiny  FKKR} &{\tiny  FLAPW} & {\tiny  LMTO} &
{\tiny  FKKR} & {\tiny FLAPW} &{\tiny  LMTO} &{\tiny  FKKR}
&{\tiny   FLAPW} &{\tiny LMTO} \\ \hline (111) &  0.67 & 0.62
&0.71& 0.57 & 0.51 & 0.55 &0.62 & 0.50 &0.61 \\ (100) &  0.87 &
0.81 &0.91& 0.73 & 0.65 & 0.65 &0.84 & 0.68 &0.90 \\ (110) &  1.33
& 1.25 &1.32& 1.11 & 1.00 & 0.95 &1.28 & 1.01 &1.32
\end{tabular}
\end{table}

\begin{table}
\caption{FKKR scalar-relativistic surface energies  given in  eV/(surface atom) 
for the seven more close packed
surfaces together with the anisotropy ratios  in parenthesis. $d$
is the distance between two successive layers and $a$ the lattice
constant. BB is the number of first-neighbor broken bonds. 
} \label{table2}
\begin{tabular}{rrrrrr}
& Cu  & Ag  & Au & $d/a$ & BB   \\  \hline (111) &
0.675 &              0.566 &       0.623 & 0.5774 & 3     \\ (100) &
{\em (1.29)} 0.874 & {\em (1.29)} 0.728 & {\em (1.35)} 0.842 &
  0.5000 & 4  \\
(110) & {\em (1.97)} 1.327 & {\em (1.97)} 1.113 & {\em (2.06)} 1.284
&      0.3535 & 6    \\
(311) & {\em (2.32)} 1.564 & {\em (2.31)} 1.309  &  {\em (2.36)}
1.468 & 0.3015 & 7 \\
 (331) & {\em (2.99)} 2.016 & {\em (2.97)} 1.680 & {\em (3.05)} 1.900 &
0.2294 & 9 \\ 
(210) & {\em (3.32)} 2.240  & {\em (3.29)} 1.864  & {\em (3.45)} 2.149 & 0.2236 & 10 \\ 
(211) & {\em (3.34)} 2.255  & {\em (3.32)} 1.877  & {\em (3.39)} 2.110&0.2041 & 10
\end{tabular}
\end{table}

\end{document}